\documentclass[12pt]{article}
\title{\bf On the Possibility of Building a Black Hole for 
Electrons in the Laboratory}
\author{  Murat \"Ozer \\
Department of Physics, College of Science,
King Saud University,\\
P. O. Box 2455, Riyadh 11451, Saudi Arabia}
\begin{document}
\maketitle

\begin{abstract}
\noindent We point out in this work that if our recently proposed unified 
description of gravitation
and electromagnetism through a symmetric metric tensor is true, then
building in the laboratory  black holes for electrons with radii 
$r_E\ge 0.5m$ in air and with much smaller radii in a vacuum should be
possible.
\end{abstract}\vspace{0.5cm}

The possibility that spacetime is curved in the presence of an 
electromagnetic field, just like it is curved in the presence of a
gravitational field, was recently studied \cite{unif} and experiments to test
this conjecture were proposed \cite{exp}. If the ideas introduced in ref.[1]
are correct, then not only do we have a unified description of gravitation
 and electromagnetism through a symmetric metric tensor $g_{\mu\nu}$,
but also new physical phenomena such as the redshift and deflection of
light in an electric and magnetic field, respectively \cite{exp}. When a 
negatively(positively) charged test particle
moves in the viscinity of a positively(negatively) charged sphere of 
radius $R$ and mass $M$, 
with an electric potential of $V(R)$ on its surface,  the line
element is given by \cite{exp}
\begin{eqnarray}
ds^2 &=&-\left(1-2\frac{GM}{c^2r}-2\frac{\mid\mp q\mid}{m}\frac{R\mid\pm 
V(R)\mid}{c^2r}\right)c^2dt^2+\nonumber\\
& &\left(1-2\frac{GM}{c^2r}-2\frac{\mid\mp q\mid}{m}\frac{R\mid\pm 
V(R)\mid}{c^2r}\right)^{-1}dr^2
+r^2d\theta^2+r^2sin^2\theta d\phi^2,
\end{eqnarray}
where $G$ is the gravitational constant, $c$ is the speed of light, $q$
is the electric charge, and $m$ is the mass of the test particle. Eq.(1)
reduces to the Schwarzschild line element \cite{sch} when $V(R)=0$. 
One noteworthy point about the solution in eq.(1) is that test
particles with different electric charge-to-mass ratios move in 
different spacetime manifolds. Inspection of eq.(1) reveals that
\begin{eqnarray}
g_{00}=-\left(1-2\frac{m_G}{r}-2\frac{m_E}{r}\right)\rightarrow 0,
\nonumber\\
g_{11}=\left(1-2\frac{m_G}{r}-2\frac{m_E}{r}\right)^{-1}\rightarrow\infty,
\end{eqnarray}
as
\begin{equation}
r\rightarrow 2(m_G+m_E),
\end{equation}
where 
\begin{equation}
m_G=\frac{GM}{c^2},\hspace{0.5cm}m_E=\frac{\mid\mp q\mid}{m}\frac{R\mid\pm 
V(R)\mid}{c^2}.
\end{equation}
Then there is an {\em event horizon} at $r_{EH}=2(m_G+m_E)$ and the
region enclosed by it will be a black hole \cite{black}, for those particles
whose electric charge-to-mass ratio is $q/m$, if $r_{EH}>R$, the radius 
of the
sphere. We wish to show in this letter that such a black hole 
for electrons can be built
in the laboratory, provided, of course, that our unified description of
gravitation and electromagnetism is true.

Consider a sphere of radius $R=0.5m$, mass M=1000kg, and charged to
a potential difference of $3\times 10^5V$ with respect to the ground,
 and with electrons in its viscinity. 
Then $m_G=7.42\times 10^{-25}m$, and $m_E=0.29m$, indicating that the
gravitational field of the sphere is negligibly weak compared to its
electric field. Thus we can write
\begin{equation}
g_{00}=-\left(1-2\frac{m_E}{r}\right),\hspace{0.5cm}
g_{11}=\left(1-2\frac{m_E}{r}\right)^{-1}.
\end{equation}
The {\em event horizon} then occurs at the {\em electrical radius}
\begin{equation}
r_E=2m_E=2\frac{e}{m}\frac{RV(R)}{c^2},
\end{equation}
which corresponds to the {\em gravitational radius} (Schwarzschild
radius) $r_S=2m_G$ in pure gravitation
\footnote{There corresponds a different electrical radius $r_E$ to each
 $q/m$ ratio in the same electric field. This is in sharp contrast with 
gravitation in which
all the  test particles have the same gravitational radius $r_G$ in
the same gravitational field.}. Here $e$ and $m$ are the charge and the mass of the electron, 
respectively. The values of the potential 
$V(R)$ that
causes $r_E$ to be larger than the radius $R$ of the sphere are then 
given, for electrons, by
\begin{equation}
V(R)=\frac{r_E}{R}\frac{mc^2}{2e},\hspace{0.5cm} r_E>R.
\end{equation}

Tabulated in Table 1 are $r_E$, $V(R)$, and the electric field strength
$V(R)/R$ on the sphere for sphees with different radii. The reason we
quote the electric field strength $V(R)/R$ is that if $V(R)/R>3\times 10^6$
, which is the {\em dielectric strength} of air, then air will begin to
conduct electricity and discharging from the sphere will occur. We see that
a black hole, for electrons,  can be built in the laboratory rather 
easily. Let some
electrons be injected continously into the viscinity of a metallic sphere
positively charged
\footnote{It is seen from Table 1 that the potential $V(R)$ must be larger
than $255,495V$. One way of obtaining such a high potential may
be connecting the sphere, through a cable, to a Van de Graaf generator 
already producing a high enough potential.}
 according to a desired electrical radius $r_E$. Electrons that are
inside the spherical surface of radius $r_E$, i.e. the black hole,
 cannot go out of this surface. 
Had the ``effective'' electric charge-to-mass ratio of the photons \cite{exp} 
were the same as that of the electrons, we would have had the possibility
of building black holes for the light too. Unfortunately, photons
should have, if at all, a different ratio, most probably 
$(q/m)_{\gamma}=\pm 1$ \cite{exp}. As a result of the nonequality of their
respective electric charge-to-mass ratios, electrons and photons
move in different geometries in the same electric field. Note, 
however, that if 
$\mid (q/m)_{\gamma}\mid$=1, the potential required to produce an electrical 
black hole for light with $r_E/R=2$ is $9\times 10^{16}V$.
It should be noted that
because of the dielectric strength, $3\times 10^6V/m$, of air small size
spheres can maintain their electric charge only in a vacuum.\\

Finally, we would like to point out that  energy considerations in 
Newtonian electricity, namely Laplace's theorem {\em `that the attractive 
force of a heavenly body could be so large that light could not flow out of
it'} \cite{lap} parallelled in electricity, yield the same expression as 
ours in eq.(6): Consider
an electron moving radially away from a positively charged sphere of
charge $Q$ and radius $R$. The electron will escape from
a distance of $r$ and reach infinity at zero speed:
\begin{equation}
\frac{1}{2}m_ev_{esc}^2-\frac{k_eeQ}{r}=0,
\end{equation}
where the gravitational potential energy of the electron has been 
neglected.
The radius $r_E$ of the spherical surface from which the electron cannot
escape is obtained by replacing $v_{esc}$ by $c$, the speed of light:
\begin{equation}
r_E=2\frac{e}{m_e}\frac{k_eQ}{c^2}.
\end{equation}
With  $k_eQ=RV(R)$, this expression for $r_E$ is the same as that obtained 
from the metric which describes the
dynamical actions of gravitation and electricity together. This
argument may be a strong indication that our metric unification of 
gravitation and 
electromagnetism may indeed be true. If gravitational black holes are a 
reality, then so must be the electrical ones according not only to our
scheme, but also to Newtonian electricity. 

In conclusion, if our scheme for a unified description of gravitation and
electromagnetism \cite{unif, exp} is true, we then have the intriguing 
possibility of building  (electrical) black holes for electrons in the 
laboratory. The study of nonrotating as well as rotating black holes for 
electrons in the laboratory may soon be a reality.  
\begin{table}
\label{Table 1}
\caption{Values of the electrical radius $r_E$, the required potential 
$V(R)$, and the electric field strength $V(R)/R$ at various values of the radius  $R$ of the sphere.}\vspace{0.5cm}
\begin{tabular}{|c|c|c|c|} \hline
$R$ & $r_E$ & $V(R)$ & $V(R)/R$  \\
(m) &  (m)  & (V)     & (V/m) \\ \cline{1-4}
0.05 & 0.05 & 255,495 & $5.11\times 10^6$\\
     & 0.10 & 510,990 & $7.66\times 10^6$\\
     & 0.15 & 766,484 & $1.53\times 10^7$\\
     & 0.20 &1,021,980& $2.04\times 10^7$\\
0.10 & 0.10 & 255,495 & $2.55\times 10^6$\\
     & 0.15 & 383,242 & $3.83\times 10^6$\\
     & 0.20 & 510,990 & $5.11\times 10^6$\\
     & 0.25 & 638.737 & $6.39\times 10^5$\\
0.25 & 0.25 & 255,495 & $1.02\times 10^6$\\
     & 0.30 & 306,594 & $1.22\times 10^6$\\
     & 0.40 & 408,792 & $1.64\times 10^6$\\
     & 0.50 & 510,990 & $2.04\times 10^6$\\
0.50 & 0.50 & 255,495 & 510,990\\
     & 0.60 & 306,594 & 587,638\\
     & 0.70 & 357,693 & 715,383\\
     & 0.80 & 408,792 & 817,583\\
     & 0.90 & 459,891 & 919,781\\
     & 1.00 & 510,990 & $1.02\times 10^6$\\
1.00 & 1.00 & 255,495 & 255,495\\
     & 1.25 & 319,369 & 319,369\\
     & 1.50 & 383,242 & 383,242\\
     & 1.75 & 447,116 & 447,116\\
     & 2.00 & 510,990 & 510,990\\
2.50 & 2.50 & 255,495 & 102,198\\
     & 2.75 & 281,044 & 112,418\\ 
     & 3.00 & 306,595 & 122,638\\
     & 3.50 & 357,693 & 143,077\\
     & 4.00 & 408,792 & 163,517\\
     & 4.50 & 459,891 & 183,956\\
     & 5.00 & 510,990 & 204,396\\
\hline
\end{tabular}
\end{table}

\end{document}